\documentclass[onecolumn,showpacs]{revtex4}

\usepackage{graphicx}
\usepackage{dcolumn}
\usepackage{bm}

\input epsf

\begin{document}

\title{\Large  Role of Modified Chaplygin Gas in Accelerated Universe}

\author{\bf Ujjal Debnath$^1$\footnote{ujjaldebnath@yahoo.com},
~Asit Banerjee$^2$\footnote{asitb@cal3.vsnl.net.in} and~Subenoy
Chakraborty$^1$\footnote{subenoyc@yahoo.co.in}}

\affiliation{ $^1$Department of Mathematics, Jadavpur University,
Calcutta-32, India.\\ $^2$Department of Physics, Jadavpur
University, Calcutta-32, India.}

\date{\today}

\begin{abstract}
In this paper we have considered a model of modified Chaplygin gas
and its role in accelerating phase of the universe. We have
assumed that the equation of state of this modified model is valid
from the radiation era to $\Lambda$CDM model. We have used
recently developed statefinder parameters in characterizing
different phase of the universe diagrammatically.
\end{abstract}

\pacs{}

\maketitle

\section{\normalsize\bf{Introduction}}

Recent observations of the luminosity of type Ia supernovae
indicate [1, 2] an accelerated expansion of the universe and lead
to the search for a new type of matter which violate the strong
energy condition $\rho+3p<0$. The matter consent responsible for
such a condition to be satisfied at a certain stage of evolution
of the universe is referred to as {\it dark energy}. There are
different candidates to play the role of the dark energy. The
type of dark energy represented by a scalar field is often called
{\it quintessence}. The transition from a universe filled with
matter to an exponentially expanding universe does not necessarily
require the presence of the scalar field as the only alternative.
In particular one can try another alternative by using an exotic
type of fluid - the so-called Chaplygin gas which obeys an
equation of state like $p=-B/\rho$ [3] where $p$ and $\rho$ are
respectively the pressure and energy density and $B$ is a positive
constant. Subsequently the above equation was modified to the form
$p=-B/\rho^{\alpha}$ with $0\le \alpha \le 1$. This model gives
the cosmological evolution from an initial dust like matter to an
asymptotic cosmological constant with an epoch that can be seen as
a mixture of a cosmological constant and a fluid obeying an
equation of state $p=\alpha\rho$. This generalized model has been
studied previously [4, 5].\\

In the paper [4], the flat Friedmann model filled with Chaplygin
fluid has been analyzed in terms of the recently proposed ``{\it
statefinder}'' parameters [6]. In fact trajectories in the
$\{s,r\}$ plane corresponding to different cosmological models
demonstrate qualitatively different behaviour. The statefinder
diagnostic along with future SNAP observations may perhaps be used
to discriminate between different dark energy models. The above
statefinder diagnostic pair are constructed from the scale factor
$a(t)$ and its derivatives upto the third order as follows:

\begin{equation}
r=\frac{\dddot{a}}{aH^{3}}~~~~\text{and}~~~~s=\frac{r-1}{3\left(q-\frac{1}{2}\right)}
\end{equation}

where $H$ and $q~\left(=-\frac{a\ddot{a}}{\dot{a}^{2}}\right)$ are
the Hubble parameter and the deceleration parameter respectively.
These parameters are dimensionless and allow us to characterize
the properties of dark energy in a model independent manner. The
parameter $r$ forms the next step in the hierarchy of geometrical
cosmological parameters after $H$ and $q$.\\

In section II, all the discussions are valid in general for $k=0,
\pm 1$, but in section III we have specifically considered the
simple case of a spatially flat universe ($k=0$), which naturally
follows from the simplest version of the inflationary scenario
and is confirmed by recent CMB experiments [7]. In our present
work we consider a more general modified Chaplygin gas obeying an
equation of state [8]

\begin{equation}
p=A\rho-\frac{B}{\rho^{\alpha}} ~~~~\text{with}~~~~ 0\le \alpha
\le 1
\end{equation}

This equation of state shows radiation era (when $A=1/3$) at one
extreme (when the scale factor $a(t)$ is vanishingly small) while
a $\Lambda$CDM model at the other extreme (when the scale factor
$a(t)$ is infinitely large). At all stages it shows a mixture.
Also in between there is also one stage when the pressure
vanishes and the matter content is equivalent to a pure dust. We
have further described this particular cosmological model from
the field theoretical point of view by introducing a scalar field
$\phi$ and a self-interacting potential $V(\phi)$ with the
effective Lagrangian

\begin{equation}
{\cal L_{\phi}}=\frac{1}{2}\dot{\phi}^{2}-V(\phi)
\end{equation}

In the paper of Gorini et al [4], it has been shown that for the
simple flat Friedmann model with Chaplygin gas can equivalently
described in terms of a homogeneous minimally coupled scalar field
$\phi$. In this case FRW equations for Chaplygin gas fits in
Barrow's scheme [9]. Following Barrow [10], Kamenshchik et al [3,
11] have obtained homogeneous scalar field $\phi(t)$ and a
potential $V(\phi)$ to describe Chaplygin cosmology. In the
section II, we have obtained the corresponding expressions for a
modified Chaplygin gas as generalization of the previous work.\\

It has been possible to express $V(\phi)$ in terms of the scale
factor $a(t)$ for arbitrary values of the constant $\alpha$ and
have explicitly shown how $V(\phi)$ varies as the scale factor
interpolates between radiation and $\Lambda$CDM stages for
different values of the constant $A$. In the next stage evolution
of the model universe has been studied in the $\{r,s\}$ plane for
the entire physically realistic history of the universe and
compared to that for either pure Chaplygin gas model or a
generalized Chaplygin gas model as given in [5].\\

\section{\normalsize\bf{Modified Chaplygin Gas in FRW Model}}

The metric of a homogeneous and isotropic universe in FRW model is

\begin{equation}
ds^{2}=dt^{2}-a^{2}(t)\left[\frac{dr^{2}}{1-kr^{2}}+r^{2}(d\theta^{2}+sin^{2}\theta
d\phi^{2})\right]
\end{equation}

where $a(t)$ is the scale factor and $k~(=0,\pm 1)$ is the
curvature scalar.\\

The Einstein field equations are

\begin{equation}
\frac{\dot{a}^{2}}{a^{2}}+\frac{k}{a^{2}}=\frac{1}{3}\rho
\end{equation}
and
\begin{equation}
\frac{\ddot{a}}{a}=-\frac{1}{6}(\rho+3p)
\end{equation}

where $\rho$ and $p$ are energy density and isotropic pressure
respectively (choosing $8\pi G=c=1$).\\

The energy conservation equation is

\begin{equation}
\dot{\rho}+3\frac{\dot{a}}{a}(\rho+p)=0
\end{equation}

Using equation (2) we have the solution of $\rho$ as

\begin{equation}
\rho=\left[\frac{B}{1+A}+\frac{C}{a^{3(1+A)(1+\alpha)}}
\right]^{\frac{1}{1+\alpha}}
\end{equation}

where $C$ is an arbitrary integration constant.\\

Due to complicated form of the equation of state (2), the scale
factor $a(t)$ can not be solved from the Einstein field equations
(5) and (6) for arbitrary $k$. However for $k=0$, the solution of
scale factor $a(t)$ has the form

\begin{equation}
a^{\frac{3(1+A)}{2}}~_{2}F_{1}[x,x,1+x,-\frac{B}{C(1+A)}a^{\frac{3(1+A)}{2x}}]=\frac{\sqrt{3}}{2}(1+A)C^{x}t
\end{equation}

where $x=\frac{1}{2(1+\alpha)}$ and $_{2}F_{1}$ is the
hypergeometric function.\\

Now for small value of scale factor $a(t)$, we have

\begin{equation}
\rho\simeq \frac{C^{\frac{1}{1+\alpha}}}{a^{3(1+A)}}~,
\end{equation}

which is very large and corresponds to the universe dominated by
an equation of state $p=A\rho$.\\

Also for large value of scale factor $a(t)$,

\begin{equation}
\rho\simeq
\left(\frac{B}{1+A}\right)^{\frac{1}{1+\alpha}}~~~~\text{and}~~~~p\simeq
-\left(\frac{B}{1+A}\right)^{\frac{1}{1+\alpha}}=-\rho
\end{equation}

which correspond to an otherwise empty universe with a
cosmological constant
$\left(\frac{B}{1+A}\right)^{\frac{1}{1+\alpha}}$.\\

For accelerating universe $q$ must be negative i.e., $\ddot{a}>0$
i.e.,
\begin{equation}
a^{3(1+\alpha)(1+A)}>\frac{C(1+A)(1+3A)}{2B}
\end{equation}

This expression shows that for small value of scale factor we have
decelerating universe while for large values of scale factor we
have accelerating universe and the transition occurs when the
scale factor has the expression $a=\left[\frac{C(1+A)(1+3A)}{2B}\right]^{\frac{1}{3(1+\alpha)(1+A)}}$.\\

Considering now the subleading terms in equation (8) at large
values of $a$, we can obtain the following expressions for the
energy density and pressure:

\begin{equation}
\rho\simeq
\left(\frac{B}{1+A}\right)^{\frac{1}{1+\alpha}}+\frac{C}{1+\alpha}\left(\frac{1+A}
{B}\right)^{\frac{1}{1+\alpha}}~a^{-3(1+\alpha)(1+A)}
\end{equation}
and
\begin{equation}
p\simeq
-\left(\frac{B}{1+A}\right)^{\frac{1}{1+\alpha}}+\frac{C}{1+\alpha}\left(\frac{1+A}
{B}\right)^{\frac{1}{1+\alpha}}\{\alpha+(1+\alpha)A\}~a^{-3(1+\alpha)(1+A)}
\end{equation}

Equations (13) and (14) describe the mixture of a cosmological
constant equal to
$\left(\frac{B}{1+A}\right)^{\frac{1}{1+\alpha}}$ with a matter
whose equation of state is given by

\begin{equation}
p=\{\alpha+(1+\alpha)A\}\rho~,
\end{equation}

which for a pure Chaplygin gas reduces to $p=\alpha\rho$.\\

Now we consider this energy density and pressure corresponding to
a scalar field $\phi$ having a self-interacting potential
$V(\phi)$. The Lagrangian of the scalar field has been given in
equation (3). The analogous energy density $\rho_{\phi}$ and
pressure $p_{\phi}$ for the scalar field are the following:

\begin{equation}
\rho_{\phi}=\frac{1}{2}\dot{\phi}^{2}+V(\phi)=\rho=\left[\frac{B}{1+A}+\frac{C}{a^{3(1+A)(1+\alpha)}}
\right]^{\frac{1}{1+\alpha}}
\end{equation}
and
\begin{equation}
p_{\phi}=\frac{1}{2}\dot{\phi}^{2}-V(\phi)=A\rho-\frac{B}{\rho^{\alpha}}
=A \left[\frac{B}{1+A}+\frac{C}{a^{3(1+A)(1+\alpha)}}
\right]^{\frac{1}{1+\alpha}}-B
\left[\frac{B}{1+A}+\frac{C}{a^{3(1+A)(1+\alpha)}}
\right]^{-\frac{\alpha}{1+\alpha}}
\end{equation}
\\
Hence for flat universe (i.e., $k=0$) we have

\begin{equation}
\dot{\phi}^{2}=(1+A)\left[\frac{B}{1+A}+\frac{C}{a^{3(1+A)(1+\alpha)}}
\right]^{\frac{1}{1+\alpha}}-B
\left[\frac{B}{1+A}+\frac{C}{a^{3(1+A)(1+\alpha)}}
\right]^{-\frac{\alpha}{1+\alpha}}
\end{equation}
and
\begin{equation}
V(\phi)=\frac{1}{2}(1-A)\left[\frac{B}{1+A}+\frac{C}{a^{3(1+A)(1+\alpha)}}
\right]^{\frac{1}{1+\alpha}}+\frac{1}{2}B
\left[\frac{B}{1+A}+\frac{C}{a^{3(1+A)(1+\alpha)}}
\right]^{-\frac{\alpha}{1+\alpha}}
\end{equation}

From equation (18) we have the relation between scale factor
$a(t)$ and scalar field $\phi$ as

\begin{equation}
\phi=\frac{2}{\sqrt{3}(1+\alpha)(1+A)}~Sinh^{-1}\left\{\sqrt{\frac{C(1+A)}{B}}\frac{1}{a^{\frac{3}{2}(1+\alpha)(1+A)}}
\right\}
\end{equation}

From equation (19) we get

\begin{eqnarray*}
V(\phi)=\frac{1}{2}(1-A)\left(\frac{B}{1+A}\right)^{\frac{1}{1+\alpha}}Cosh^{\frac{2}{1+\alpha}}
\left\{\frac{\sqrt{3}\sqrt{1+A}(1+\alpha)}{2}\phi\right\}
\end{eqnarray*}

\begin{equation}
+\frac{1}{2}B\left(\frac{B}{1+A}\right)^{-\frac{\alpha}{1+\alpha}}Cosh^{-\frac{2\alpha}{1+\alpha}}
\left\{\frac{\sqrt{3}\sqrt{1+A}(1+\alpha)}{2}\phi\right\}
\end{equation}

Since $2V(\phi)=\rho-p=(1-A)\rho+\frac{B}{\rho^{\alpha}}$. So
when $\rho\rightarrow\infty$ i.e., when $a\rightarrow 0$ the
situations are completely different for $A\ne 1$ and for $A=1$.
In the second case $V(\phi)\rightarrow 0$ as
$\rho\rightarrow\infty$, but in the first case
$V(\phi)\rightarrow\infty$ as $\rho\rightarrow\infty$. This is
why we get qualitatively two different pictures in two cases - as
reflected in the figures 2 and 5. In the other limit
$a\rightarrow\infty$, $V(\phi)\rightarrow
\left(\frac{B}{1+A}\right)^{\frac{1}{1+\alpha}}$ for the above
two cases.  \\

For example, $A=\frac{1}{3},~B=C=\alpha=1$, $V(\phi)\rightarrow
\frac{\sqrt{3}}{2}$ as $a\rightarrow\infty$\\

~~~~~~and ~~~~~~$A=1,~B=C=\alpha=1$,
$V(\phi)\rightarrow\frac{1}{\sqrt{2}}$ as $a\rightarrow\infty$.\\

\begin{figure}
\includegraphics[height=1.7in]{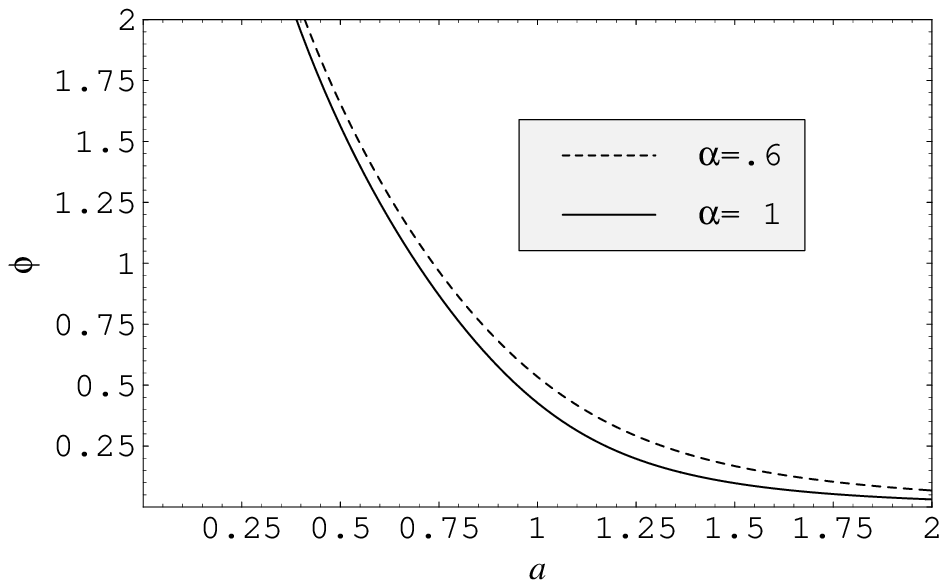}~~~
\includegraphics[height=1.7in]{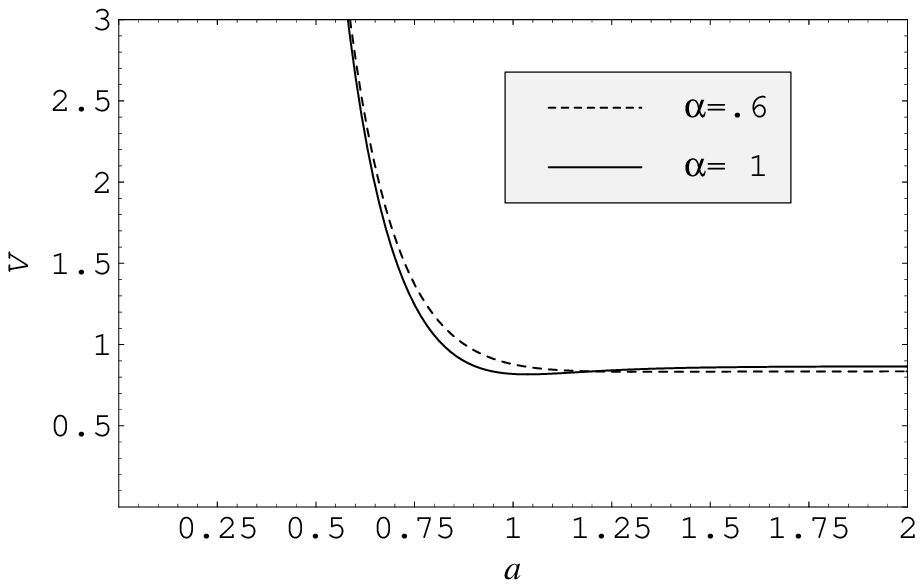}\\
\vspace{1mm}
Fig.1~~~~~~~~~~~~~~~~~~~~~~~~~~~~~~~~~~~~~~~~~~~~~~~~~~~~~Fig.2\\
\vspace{5mm}
\includegraphics[height=1.7in]{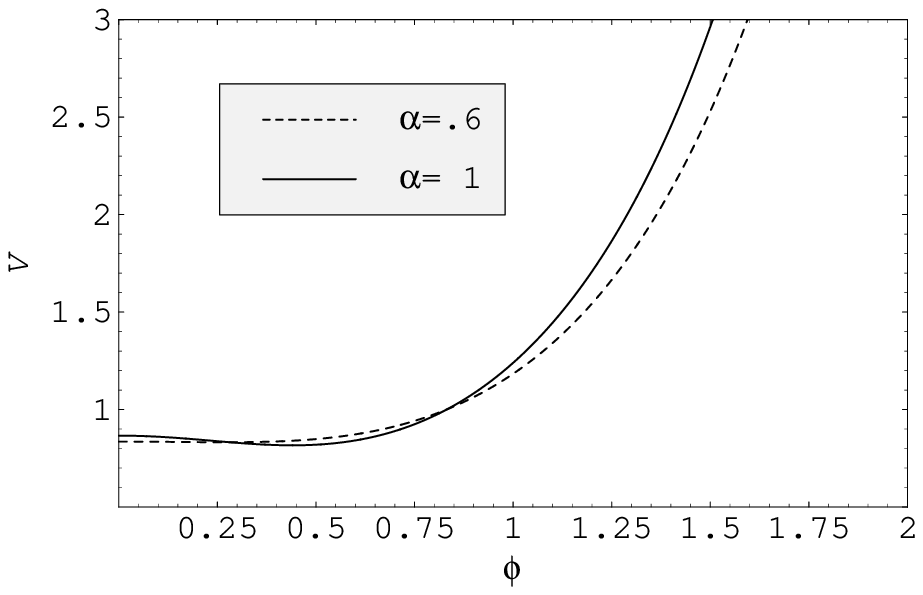}\\
Fig.3

\vspace{7mm} Figs. 1 - 3 shows variation of $\phi$ and $V$ against
$a$ and $\phi$ for $A=1/3$ and $\alpha=~ 0.6,~ 1$.\hspace{2cm}
\vspace{6mm}

\includegraphics[height=1.7in]{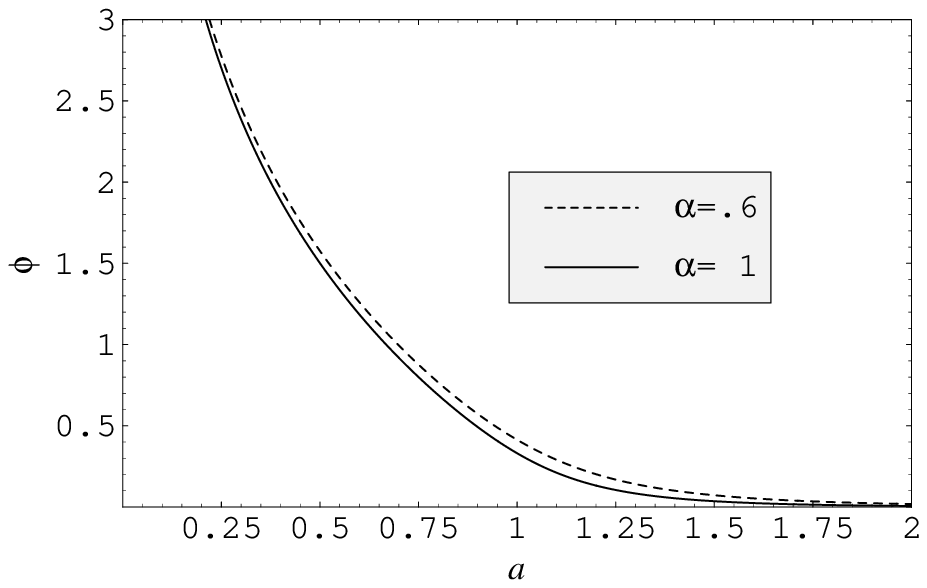}~~~
\includegraphics[height=1.7in]{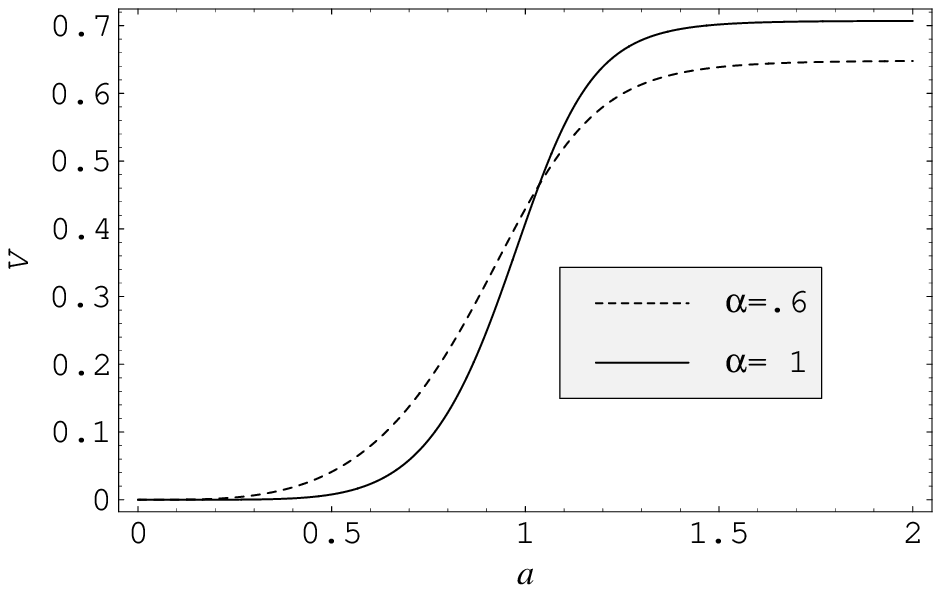}\\
\vspace{1mm}
Fig.4~~~~~~~~~~~~~~~~~~~~~~~~~~~~~~~~~~~~~~~~~~~~~~~~~~~~Fig.5\\
\vspace{5mm}
\includegraphics[height=1.7in]{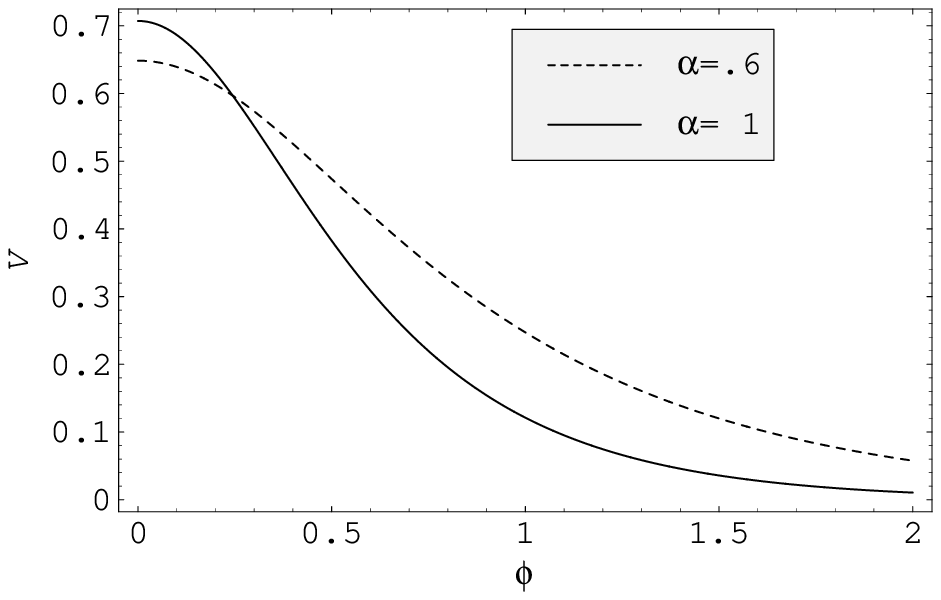}\\
Fig.6

\vspace{5mm} Figs. 4 - 6 shows variation of $\phi$ and $V$ against
$a$ and $\phi$ for $A=1$ and $\alpha=~0.6,~ 1$.\hspace{2cm}
\vspace{6mm}

\end{figure}

The graphical representation of $\phi$ against $a$ and $V(\phi)$
against $a$ and $\phi$ respectively have been shown in figures 1 -
3 for $A=1/3$ and figures 4 - 6 for $A=1$. From figures 1, 4 we
have seen that scalar field $\phi$ sharply falls when scale factor
$a(t)$ increases, both for $A=1/3$ and $A=1$. In figure 2, we see
that potential function $V(\phi)$ sharply decreases from extremely
large value and thereafter it increases to a fixed value for
$A=1/3$. On the other hand for $A=1$, the nature of $V(\phi)$
against $a$ is as stated in the following. For small $a$,
$V(\phi)$ varies insignificantly from zero value and then
increases sharply to a constant value. In figure 3, $V(\phi)$
decreases slightly and then increases to infinitely large value
for $A=1/3$ while for $A=1$, in figure 6, $V(\phi)$ decreases to
zero asymptotically starting from a extremely large value. So
$V(\phi)$ demonstrates completely different behaviour for two
different values of $A=1$ and $A=1/3$. Further the figures show
how $V(\phi)$ varies with $\phi$ with different values chosen for
$\alpha$.\\

\section{\normalsize\bf{The Role of StateFinder parameters in FRW Universe}}

The statefinder parameters have been defined in equation (1).
Trajectories in the $\{r,s\}$ plane corresponding to different
cosmological models, for example $\Lambda$CDM model diagrams
correspond to the fixed point $s=0,~r=1$.\\

For Friedmann model with flat universe (i.e., $k=0$),

\begin{equation}
H^{2}=\frac{\dot{a}^{2}}{a^{2}}=\frac{1}{3}\rho
\end{equation}
and
\begin{equation}
q=-\frac{\ddot{a}}{aH^{2}}=\frac{1}{2}+\frac{3}{2}\frac{p}{\rho}
\end{equation}

So from equation (1) we get

\begin{equation}
r=1+\frac{9}{2}\left(1+\frac{p}{\rho}\right)\frac{\partial
p}{\partial\rho}~,~~~~~s=\left(1+\frac{\rho}{p}\right)\frac{\partial
p}{\partial\rho}
\end{equation}

Thus we get the ratio between $p$ and $\rho$:

\begin{equation}
\frac{p}{\rho}=\frac{2(r-1)}{9s}
\end{equation}

For modified Chaplygin gas, velocity of sound can be written as

\begin{equation}
v_{s}^{2}=\frac{\partial p}{\partial\rho}=A(1+\alpha)-\frac{\alpha
p}{\rho}
\end{equation}

From equations (24) and (25) we get the relation between $r$ and
$s$:

\begin{equation}
18(r-1)s^{2}+18\alpha s(r-1)+4\alpha
(r-1)^{2}=9sA(1+\alpha)(2r+9s-2)
\end{equation}

In the $\{r,s\}$ plane the above equation has only one asymptote
parallel to $s$-axis namely, $r=1+\frac{9}{2}A(1+\alpha)$ and the
asymptote intersects the curve in only one point
$\left(1+\frac{9}{2}A(1+\alpha),
\frac{\alpha(1+\alpha)A}{(1+\alpha)A-\alpha}\right)$.\\

Figure 7 shows the variation of $s$ with the variation of $r$ for
$A=1/3$ and for $\alpha=~0.6,~1$. The portion of the curve on the
positive side of $s$ which is physically admissible is only the
values of $r$ greater than $\left\{1+\frac{9}{2}A(1+A)\right\}$.
The part of the curve between $r=1$ and $r=1+\frac{9}{2}A(1+A)$
with positive value of $s$ is not admissible (we have not shown
that part in the figure 7) because for the Chaplygin gas under
consideration we face a situation, where the magnitude of the
constant $B$ becomes negative. It can be shown
in the following way.\\

\begin{figure}
\includegraphics[height=2.7in]{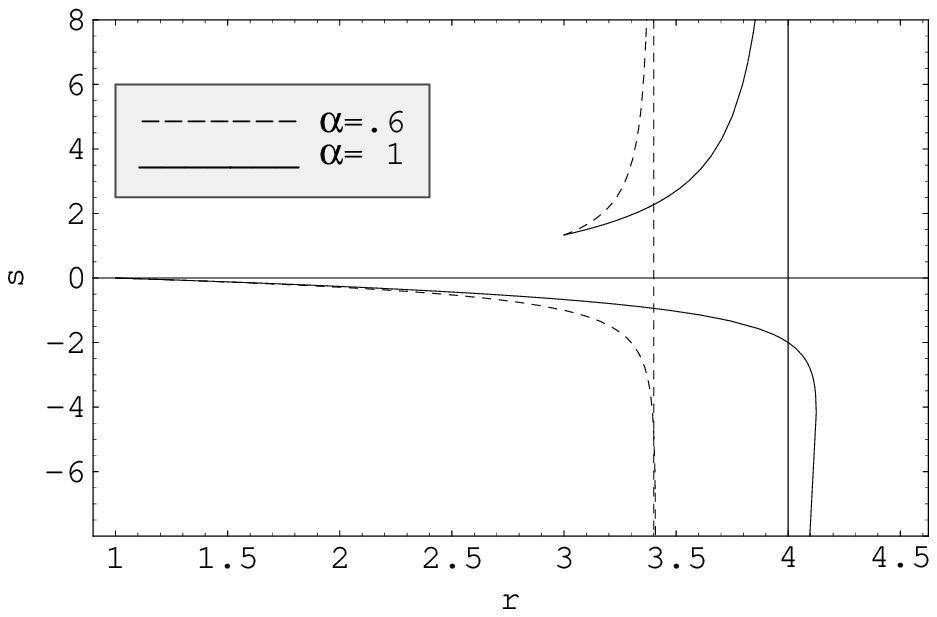}\\
\vspace{1mm} Fig.7\\

\vspace{5mm} Fig. 7 shows the variation of $s$
 against $r$ for different values of $\alpha$ (= ~0.6,~1) and for $A=1/3$.\hspace{2cm} \vspace{4mm}

\end{figure}

Since we consider the range of $A$ as $0<A<1/2$, we immediately
get the inequality relation

\begin{equation}
r-1-9A<0
\end{equation}

for all points lying in the part of the curve between $r=1$ and
$r=1+\frac{9}{2}A(1+A)$ as stated above. In this range positive
values of the parameter $s$ is explicitly given by

\begin{equation}
s=\frac{1}{2}\frac{(r-1)(1-2A)}{9A+(1-r)}\left[1+\sqrt{1+\frac{8}{9}\frac{(9A+1-r)}{(1-2A)^{2}}}~
\right]
\end{equation}

The above expression for $s$ is obtained by solving equation (27).
One must note that the other root containing negative sign before
the square root symbol is to be left out since we consider only
the positive magnitude of $s$.\\

Again from (25) and (2), we get
$$
\frac{p}{\rho}=A-\frac{B}{\rho^{\alpha+1}}=\frac{2(r-1)}{9s}
$$
which in view of (29) leads to the relation

\begin{equation}
\frac{B}{\rho^{\alpha+1}}=\frac{1}{2}-\frac{(1-2A)}{2}\sqrt{1+\frac{8}{9}\frac{(9A+1-r)}{(1-2A)^{2}}}
\end{equation}

One can show by a simple algebraic exercise that the right hand
side of (30) is clearly negative for all points corresponding to
$r<1+\frac{9}{2}A(1+A)$ and hence $B$ becomes negative, which is
unacceptable in view of the Chaplygin gas model we are
considering. This upper limit for $r$ corresponds to $r=3$ for
$A=1/3$ that is when we choose our universe to be a
mixture of radiation and Chaplygin gas.\\

Thus the curve in the positive side of $s$ starts from radiation
era and goes asymptotically to the dust model. But the portion in
the negative side of $s$ represents the evolution from dust state
($s=-\infty$) to the $\Lambda$CDM model ($s=0$). Thus the total
curve represents the evolution of the universe starting from the
radiation era to the $\Lambda$CDM model.\\

\section{\normalsize\bf{Discussions}}

In this work, we have presented a model for modified Chaplygin
gas. In this model we are able to describe the universe from the
radiation era ($A=1/3$ and $\rho$ is very large) to $\Lambda$CDM
model ($\rho$ is small constant). So compare to Chaplygin gas
model, the present model describe universe to a large extent. Also
if we put $A=0$ with $\alpha=1$, then we can recover the results
of the Chaplygin gas model. In figure 7, for $\{r,s\}$ diagram the
portion of the curve for $s>0$ between $r=1$ to
$r=1+\frac{9}{2}A(1+A)$ is not describable by the modified
Chaplygin gas under consideration. For example, if we choose $r=1.03,~A=1/3$ then
from the curve $s=0.01$ which corresponds to $q=3/2$ and hence we
have from the equation of state, $B<0$ which is not valid for the
specific Chaplygin gas model considered here. At the large
value of the scale factor we must have some stage where the
pressure becomes negative and hence $B$ has to be chosen
positive. It follows therefore that a portion of the curve
as mentioned above should not remain valid. \\\\

{\bf Acknowledgement:}\\

The authors are grateful to the referees' for their valuable suggestions.
They also thank IUCAA for worm hospitality where the
major part of the work was done. One of the authors (U.D) is
thankful to CSIR (Govt. of India) for
awarding a Senior Research Fellowship.\\

{\bf References:}\\
\\
$[1]$  N. A. Bachall, J. P. Ostriker, S. Perlmutter and P. J.
Steinhardt, {\it Science} {\bf 284} 1481 (1999).\\
$[2]$ S. J. Perlmutter et al, {\it Astrophys. J.} {\bf 517} 565
(1999).\\
$[3]$ A. Kamenshchik, U. Moschella and V. Pasquier, {\it Phys.
Lett. B} {\bf 511} 265 (2001); V. Gorini, A. Kamenshchik, U.
Moschella and V. Pasquier, {\it gr-qc}/0403062.\\
$[4]$ V. Gorini, A. Kamenshchik and U. Moschella, {\it Phys. Rev.
D} {\bf 67} 063509 (2003); U. Alam, V. Sahni , T. D. Saini and
A.A. Starobinsky, {\it Mon. Not. Roy. Astron. Soc.} {\bf 344}, 1057 (2003).\\
$[5]$ M. C. Bento, O. Bertolami and A. A. Sen, {\it Phys. Rev. D}
{66} 043507 (2002).\\
$[6]$ V. Sahni, T. D. Saini, A. A. Starobinsky and U. Alam, {\it
JETP Lett.} {\bf 77} 201 (2003).\\
$[7]$ A. Benoit, P. Ade, A. Amblard et al, {\it Astron.
Astrophys.}
{\bf 399} L25 (2003).\\
$[8]$ H. B. Benaoum, {\it hep-th}/0205140.\\
$[9]$ J. D. Barrow, {\it Nucl. Phys. B} {\bf 310} 743 (1988).\\
$[10]$ J. D. Barrow, {\it Phys.Lett. B} {\bf 235} 40 (1990).\\
$[11]$ V. Gorini, A. Yu. Kamenshchik, U. Moschella, V. Pasquier,
{\it Phys.Rev. D} {\bf 69} 123512 (2004).\\

\end{document}